\PassOptionsToPackage{dvipsnames}{xcolor}
\documentclass[twocolumn,twocolappendix]{aastex631}
\usepackage{graphicx}
\usepackage[dvipsnames]{xcolor}
\usepackage{multirow}
\usepackage{amsmath}
\usepackage{amssymb}
\usepackage{mathrsfs}
\let\tablenum\relax
\usepackage{siunitx}
\DeclareSIUnit\angstrom{\text {Å}}

\newcommand{\kms}{\ensuremath{\mathrm{km~s}^{-1}}}
\newcommand{\mbh}{\ensuremath{M_\mathrm{BH}}}
\newcommand{\ml}{\ensuremath{M^*/L}}
\newcommand{\Matern}{\text{Mat\'ern}}
\newcommand{\nsigma}[1]{$#1\text{-}\sigma$}
\newcommand{\msun}{\ensuremath{M_{\odot}}}
\newcommand{\mgb}{Mg~{\sc i}~$b$}
\newcommand{\vect}[1]{\boldsymbol{#1}}

\def\aap{Astron. \& Astrophys.}                
\def\apjl{Astrophys. J. Lett.}
\def\apjs{Astrophys. J. Suppl. Ser.}

\begin{document}

\title{Keck Integral-Field Spectroscopy of M87 Reveals an Intrinsically Triaxial Galaxy and a Revised Black Hole Mass}

\correspondingauthor{Emily Liepold}\email{emilyliepold@berkeley.edu}

\author
[0000-0002-7703-7077]
{Emily R. Liepold}
\affiliation{Department of Physics, University of California, Berkeley, California 94720, USA.}

\author
[0000-0002-4430-102X]
{Chung-Pei Ma}
\affiliation{Department of Physics, University of California, Berkeley, California 94720, USA.}
\affiliation{Department of Astronomy, University of California, Berkeley, California 94720, USA.}

\author
[0000-0002-1881-5908]
{Jonelle L. Walsh}
\affiliation{George P. and Cynthia Woods Mitchell Institute for Fundamental Physics and Astronomy, and Department of Physics and Astronomy, Texas A\&M University, College Station, TX 77843, USA.}



\begin{abstract}

The three-dimensional intrinsic shape of a galaxy and the mass of the central supermassive black hole provide key insight into the galaxy's growth history over cosmic time. Standard assumptions of a spherical or axisymmetric shape can be simplistic and can bias the black hole mass inferred from the motions of stars within a galaxy. Here we present spatially-resolved stellar kinematics of M87 over a two-dimensional $250\arcsec \times 300\arcsec$ contiguous field covering a radial range of 50 pc\textendash12 kpc from integral-field spectroscopic observations at the Keck II Telescope. From about 5 kpc and outward, we detect a prominent 25 \kms\ rotational pattern, in which the kinematic axis (connecting the maximal receding and approaching velocities) is $40^\circ$ misaligned with the photometric major axis of M87. The rotational amplitude and misalignment angle both decrease in the inner 5 kpc. Such misaligned and twisted velocity fields are a hallmark of triaxiality, indicating that M87 is not an axisymmetrically shaped galaxy. Triaxial Schwarzschild orbit modeling with more than 4000 observational constraints enabled us to determine simultaneously the shape and mass parameters. The models incorporate a radially declining profile for the stellar mass-to-light ratio suggested by stellar population studies. We find that M87 is strongly triaxial, with ratios of $p=0.845$ for the middle-to-long principal axes and $q=0.722$ for the short-to-long principal axes, and determine the black hole mass to be $(5.37^{+0.37}_{-0.25}\pm 0.22)\times 10^9 M_\odot$, where the second error indicates the systematic uncertainty associated with the distance to M87.

\end{abstract}

\keywords{Galaxy dynamics,  Galaxy masses,  Supermassive black holes,  Early-type galaxies,  Galaxies,  Galaxy dark matter halos,  Galaxy evolution, Galaxy kinematics, }

\section{Introduction} \label{sec:intro}

Some of the earliest dynamical evidence for the presence of a supermassive black hole (SMBH) came from M87 \citep{sargentetal1978}.  A bright asymmetric ring of radio emission around the M87 SMBH was imaged in 2019 \citep{EHT2019a}. The black hole mass (\mbh) inferred from the ring features is consistent with the value determined from stellar dynamics based on axisymmetric orbit modeling \citep{Gebhardtetal2011}, but it is nearly twice the mass inferred from dynamics of a gas disk around the hole \citep{Walshetal2013}.

M87 is classified as an elliptical galaxy based on the two-dimensional shape of the stellar light projected on the sky. However, its three-dimensional intrinsic shape has never been determined. A galaxy's intrinsic shape is a fundamental property that encodes the galaxy's past merger history and provides information about the mass ratios of the progenitor galaxies, the merger orbital parameters, gas fractions, and fraction of stars formed ex-situ. Whether a galaxy is intrinsically spherical, axisymmetric, or triaxial also impacts dynamical determinations of its SMBH mass and stellar mass, as well as any mass reconstructions based on the method of gravitational lensing.  

\begin{figure*}
\centering
\includegraphics{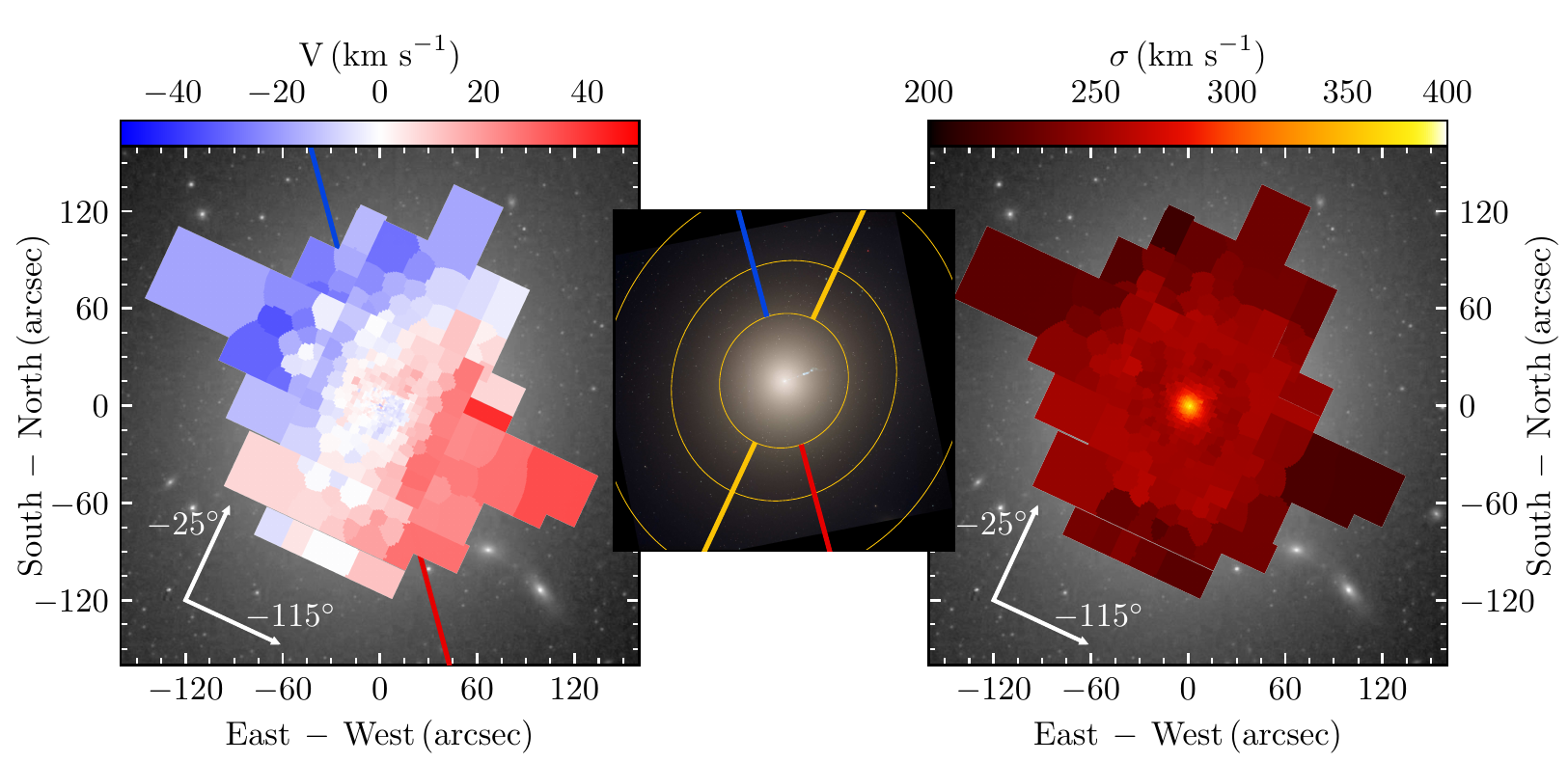}
\caption{
(Left and Right) Stellar kinematic portraits of M87 from Keck KCWI spectra in 461 spatial bins. The line-of-sight velocities (left) and velocity dispersions (right) of stars in M87 are shown over a 250\arcsec\ by 300\arcsec\ ﬁeld of view centered at the galaxy’s nucleus. The systemic velocity of M87 has been removed in the left panel. North is up and east is to the left. The two orthogonal white arrows indicate the orientations of the photometric major axis (PA of $-25^\circ$) and minor axis (PA of $-115^\circ$), as determined from the mean position angle of the galaxy’s major axis between a radius of 50\arcsec\ and 250\arcsec\ in photometric data \citep{kormendyetal2009}. The red and blue lines in the left panel mark the measured kinematic axis (PA of $-165^\circ$) outward of 70\arcsec\ (see Figure~\ref{fig:velocity}). (Middle) HST composite image of the central 200\arcsec\ by 200\arcsec\ FOV of M87, illustrating the misalignment of the photometric major axis (yellow) and kinematic axis (red-blue) beyond 50\arcsec\ along with sample isophotes of the stellar light distribution (yellow contours).
}
\label{fig:kin_map}
\end{figure*}

Thus far, almost all information about galaxy intrinsic shapes has been inferred statistically by inverting distributions of observed galaxy properties \citep{Franxetal1991, Weijmansetal2014, Foster2017, Eneetal2018, Li2018_apjl}. Here we use the Keck Cosmic Web Imager (KCWI; \citealt{Morrisseyetal2018}) on the 10 m Keck II telescope to obtain a spatially-resolved two-dimensional map of the stellar kinematics of M87 over a $250'' \times 300''$ field of view.  The resulting kinematics span a radial range of ${\sim}0\farcs6$\textendash$150''$, corresponding to a physical range of 50 pc\textendash12 kpc at a distance of $16.8\pm 0.7$ Mpc to M87, the value adopted in \citet{EHT2019f} and in this work (an angular size of 1\arcsec\ corresponds to a physical length of $81.1\pm3.3$ pc). We perform triaxial Schwarzschild orbit modeling using the detailed stellar kinematic measurements as constraints to determine M87's shape and mass parameters.  Our models include a radially declining profile for the stellar mass-to-light ratio (\ml) inferred from stellar population measurements \citep{Sarzietal2018}.

\section{Keck observations of M87} \label{sec:observations}

We observed M87 with Keck KCWI in May 2020, May 2021, March 2022, and April 2022. With the large slicer and BL grating of the integral-field unit (IFU), we obtained spectra between 3500 and 5600 \AA\ at 62 pointings, which provide contiguous two-dimensional spatial coverage of the nucleus and the outer parts of M87 (Figure~\ref{fig:kin_map}). The data span about 20 kpc (250\arcsec) across the photometric major axis ($-25^\circ$ east of north) and about 24 kpc (300\arcsec) across the photometric minor axis ($-115^\circ$ east of north).

We co-add spectra from individual KCWI spaxels to reach high signal-to-noise ratios (S/Ns), forming 461 spatial bins. Within each spatial aperture, we measure the line-of-sight stellar velocity distributions (LOSVDs) from the shapes of the absorption lines. Further details about the observations, data reduction procedures, spectral fitting processes, and stellar kinematic determination are provided in Appendices \ref{appendix:data_analysis} and \ref{appendix:kin_measurement}.

\section{Stellar Kinematic Maps}

\subsection{Misalignment between kinematic and photometric axes} \label{sec:misalignment}

The KCWI map for the line-of-sight velocity $V$ (left panel of Figure~\ref{fig:kin_map}) shows a prominent rotational pattern at large radii, in which the northeast side of the galaxy is blueshifted and the southwest side is redshifted. The kinematic axis that connects the maximal receding and approaching velocities, however, is not aligned with the photometric major axis, as it would be for an axisymmetric rotating galaxy.

\begin{figure*}
\centering
\includegraphics{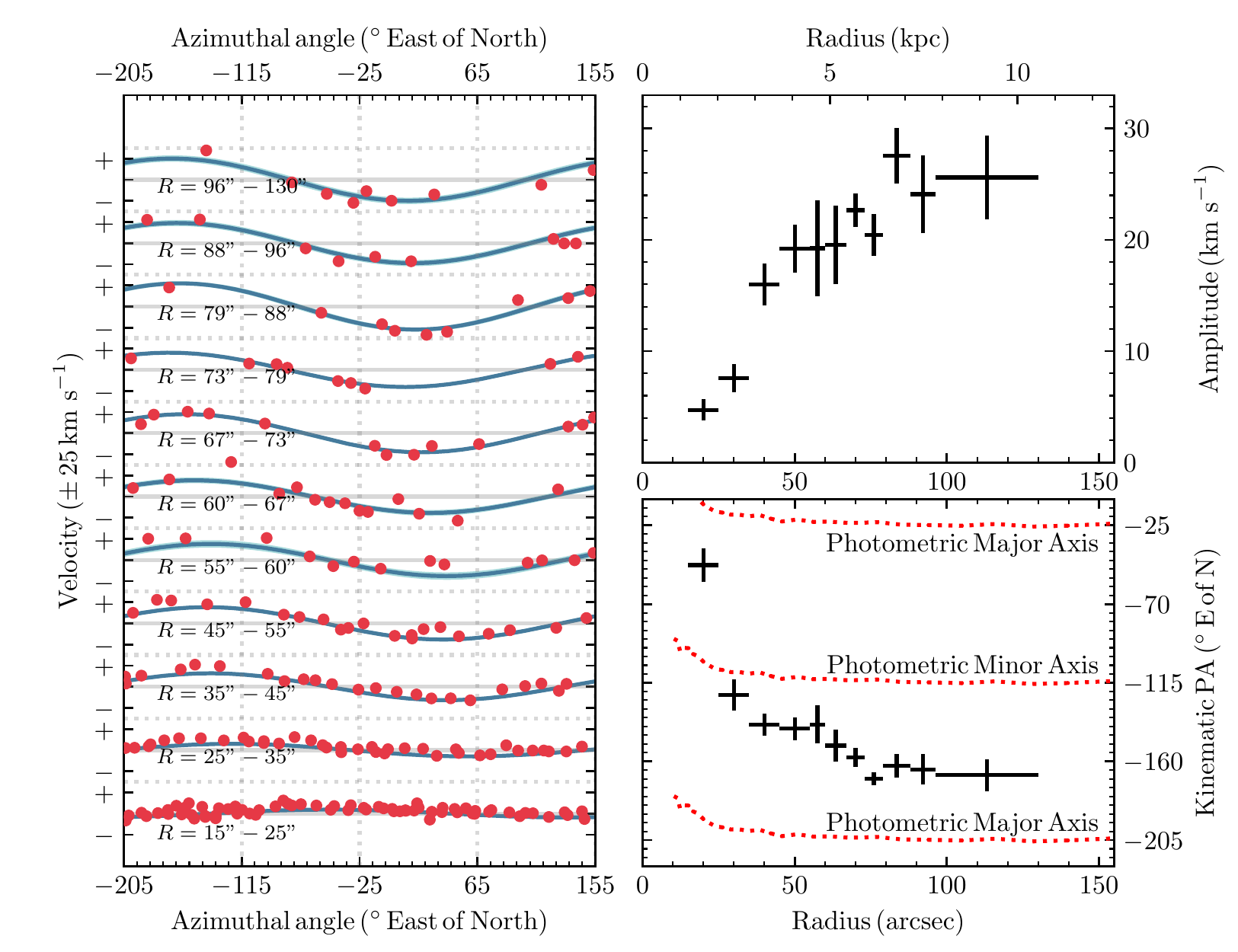}
\caption{
Line-of-sight stellar velocities as a function of projected radius and azimuthal angle on the sky.
(Left) Line-of-sight velocity as a function of azimuthal angle on the sky for 11 radial shells spanning $R=15$\arcsec\textendash130\arcsec. The velocities in each shell (red data points) are well fit (blue) by a sinusoidal function of the form $V(R,\Theta)= V_1(R) \cos\left[\Theta-\Theta_0(R)\right]$.
(Upper right) The amplitude of rotation, $V_1(R)$, increases with radius and reaches 25 \kms\ around 6 kpc.
(Lower right) The phase of the velocity function, $\Theta_0$, measures the orientation of the kinematic axis and varies significantly with radius. It plateaus to $-165^\circ$ beyond 6 kpc, indicating a $40^\circ$ misalignment between the kinematic axis and the photometric major axis (red dashed curves; \citealt{kormendyetal2009}) in M87.
}
\label{fig:velocity}
\end{figure*}

To quantify the amplitude and axis of rotation, we model the velocity field as a cosine function, with $V(R,\Theta) = V_1(R) \cos\left[\Theta-\Theta_0(R)\right]$, where $R$ is the projected radius from the galaxy's center and $\Theta$ is the azimuthal angle on the sky. The model parameters $V_1(R)$ and $\Theta_0(R)$ are the the amplitude of rotation and the position angle (PA) of the kinematic axis at radius $R$, respectively. With increasing radius, the velocity curve shows a systematic shift in phase and an increase in rotational amplitude (Figure~\ref{fig:velocity}). Within a radius of 3 kpc, the PA of the kinematic axis changes rapidly clockwise with radius (lower right panel of Figure~\ref{fig:velocity}), representing the kinematically distinct core mapped out by the Multi Unit Spectroscopic Explorer (MUSE) on the Very Large Telescope \citep{Emsellemetal2014}. Beyond 3 kpc, where the MUSE data end (at about 35\arcsec), we find that the PA of the kinematic axis continues to change clockwise and crosses the PA of the photometric minor axis, plateauing at $-165^\circ$ between 6 and 12 kpc.  Hence, there is a $40^\circ$ misalignment between the stellar kinematic axis and photometric major axis in M87.

\subsection{Stellar velocity dispersion}

\begin{figure}
\centering
\includegraphics[width=3.25in]{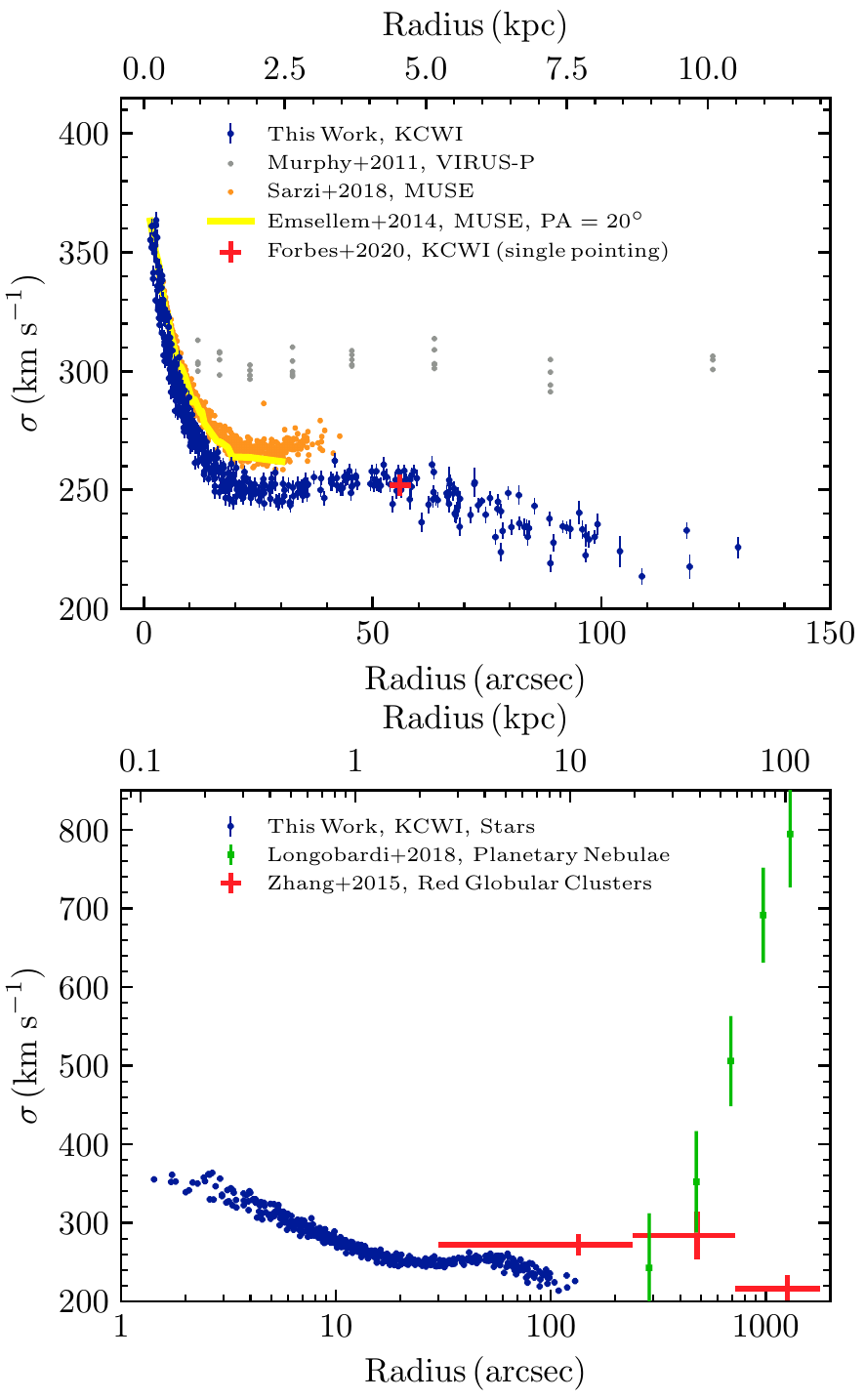}
\caption{
Radial profile of stellar velocity dispersion of M87 in the inner 10 kpc (on a linear scale; top)
and out to 100 kpc (on a logarithmic scale; bottom). All 461 KCWI bins are shown (blue) but many overlap.
(Top) The KCWI values within 1 kpc agree well with those from MUSE on the Very Large Telescope (\citealt{Emsellemetal2014,Sarzietal2018}; yellow and orange respectively), while the MUSE values are $10$\textendash$20$ \kms\ larger than KCWI between 1 and 3 kpc. At 4.5 kpc, our KCWI measurements match the single data point (red) from an independent KCWI observation \citep{Forbesetal2020}. The VIRUS-P values (\citealt{Murphyetal2011}; grey), which were used in the axisymmetric stellar-dynamical measurement of the M87 black hole \citep{Gebhardtetal2011}, are $30$\textendash$50$ \kms\ higher than all other measurements. \citet{Murphyetal2011} had noted a similar offset between their values and earlier IFU measurements \citep{Emsellemetal2004} in the inner 2 kpc.
(Bottom) Red globular clusters have similar $\sigma$ (red) as stars and appear to belong to M87's stellar halo \citep{Zhangetal2015}, whereas the intra-cluster component of planetary nebulae have sharply rising $\sigma$ (\citealt{Longobardietal2018}; green).
}
\label{fig:dispersion}
\end{figure}

The KCWI map (right panel of Figure~\ref{fig:kin_map}) and radial profile (Figure~\ref{fig:dispersion}) of the stellar velocity dispersion $\sigma$ exhibit several features. Towards the center of M87, $\sigma$ increases rapidly from 250 \kms\ at a radius of 2 kpc to 370 \kms\ at 100 pc from the nucleus. This is a clear signature of the gravitational influence of the central black hole on the motions of the stars in its vicinity. The velocity dispersion stays at about 250 \kms\ between 2 and 5 kpc and then shows a gentle 10\% decline between 5 kpc and the outermost reach of our data at 12 kpc. The stellar $\sigma$ at the edge of our field connects smoothly to the latest determinations of the velocity dispersions of discrete dynamical tracers (lower panel of Figure~\ref{fig:dispersion}) such as red globular clusters and planetary nebulae in the outer parts of M87 \citep{Zhangetal2015, Longobardietal2018}. Beyond about 10 kpc, sub-populations of planetary nebulae have been reported to have distinct kinematics \citep{Longobardietal2018}: $\sigma$ of ``intra-cluster'' planetary nebulae rises to 800 \kms\ at 100 kpc, whereas those in the galaxy halo component have a relatively flat $\sigma$ profile out to 100 kpc, similar to that of the red population of globular clusters \citep{Coteetal2001,Straderetal2011,Zhangetal2015}.

\section{Determination of mass and shape parameters from triaxial Schwarzschild modeling} \label{sec:orbit_modelling}

We use the full LOSVDs from Keck KCWI, along with photometric observations from the Hubble Space Telescope (HST) and ground-based telescopes \citep{kormendyetal2009}, to measure M87's mass distribution and intrinsic shape.  We perform triaxial Schwarzschild orbit modeling with the \texttt{TriOS} code \citep{Quennevilleetal2021,Quennevilleetal2022} based on an earlier code \citep{vandenBoschetal2008}, and use more than 4000 observational constraints to simultaneously determine six parameters: \mbh, \ml, dark matter content, and the three-dimensional intrinsic shape. As described below, we implement a new capability in the code to model spatial variations in \ml\ and use a radially declining \ml\ profile that closely approximates the variation inferred from stellar population and dynamics studies of M87 \citep{OldhamAuger2018, Sarzietal2018}.

\subsection{Galaxy model and orbit sampling}

Each galaxy model has three mass components: a central SMBH, stars, and a dark matter halo. The three-dimensional stellar density in the TriOS code is represented as a sum of multiple Gaussian functions of differing widths and axial ratios. To determine these functions, we first fit a two-dimensional Multi-Gaussian Expansion (MGE; \citealt{Cappellari2002}) to the surface brightness distribution of M87 (see Appendix~\ref{appendix:mge}). Each MGE component is allowed an independent flattening parameter ($q'$ in Table \ref{tab:mge}) to model any radially changing ellipticity observed on the sky.  

For a given set of three angles, $\theta$, $\phi$, and $\psi$, that relate the intrinsic and projected coordinate systems of a galaxy \citep{Binney1985}, we deproject each MGE component, multiply by a radially varying \ml\ (see below),  and add the deprojected Gaussians to obtain the three-dimensional stellar density. Each deprojected MGE component can have its own axis ratios $p$, $q$, and $u$, where $p=b/a$ is the intrinsic middle-to-long axis ratio, $q=c/a$ is the intrinsic short-to-long axis ratio, and $u$ is the apparent-to-intrinsic long axis ratio.  When the best-fit $p, q$, and $u$ are quoted below, each value is luminosity averaged over the MGE components. Further details of the relations between the apparent and intrinsic shape parameters and the deprojections can be found in Section~2 of \cite{Quennevilleetal2022}.

The \ml\ we use to obtain the stellar density varies radially, following a logistic curve given by
\begin{equation}
\label{eqn:MLgradient}
\frac{M^*}{L}(R) = \left(\frac{M^*}{L}\right)_{\rm outer} \left[ \frac{\delta + (R/R_0)^k}{1 + (R/R_0)^k}\right] \,,
\end{equation}
where $\delta$ is the ratio of the inner and outer \ml, and $R_0$ and $k$ parameterize the location and sharpness of the transition. We choose $\delta = 2.5$, $R_0 = 10''$, and $k = 2$, which together well approximate (Figure~\ref{fig:ML_gradient}) the spatial \ml\ profile of M87 determined from \cite{Sarzietal2018}. We leave the overall normalization\textemdash the outer \ml\textemdash as a free parameter. A similar form as Equation~(\ref{eqn:MLgradient}) was used in an axisymmetric Jeans dynamical study of M87 globular cluster and stellar kinematics data \citep{OldhamAuger2018}. We implement this spatial variation in our models by choosing distinct \ml\ ratios for each component of the MGE such that the profile is reproduced.

The dark matter halo is described by a generalized Navarro-Frenk-White density profile \citep{Navarroetal1996}
\begin{equation}
\rho(r) = \frac{\rho_0}{(r/r_s)^{\gamma}(1 + r / r_s)^{3 - \gamma}} \,,
\end{equation}
where $\rho_0$ is the density scale factor and $r_s$ is the scale radius. This form of the dark matter halo is used by \citet{Lietal2020} when fitting axisymmetric Jeans models to M87 globular cluster and stellar kinematics data. They determine that $r_s = 15.7_{-2.0}^{+2.3}$ kpc for the cored $\gamma=0$ model but find no significant preference for $\gamma = 0$ over $\gamma = 1$. \citet{OldhamAuger2016}, on the other hand, find a strong preference for flat cores with $\gamma \lesssim 0.13$. We have tested models with $\gamma = 0$, 0.5, and 1, and find that the models with a $\gamma=0$ halo are a better description of the data, with the goodness of fit ($\chi^2$) lower by at least 100. We therefore adopt the flat core, $\gamma=0$ dark matter halo. Since the KCWI stellar kinematics extend to a projected radius of 12 kpc, we expect $r_s$ and $\rho_0$ to be quite degenerate; we choose to fix $r_s = 15$ kpc and keep $\rho_0$ as a free parameter in the models.

For each galaxy model, we compute the trajectories of a library of around 500,000 stellar orbits that sample 120 values of energy, 54 and 27 values of the second integral of motion for the loop and box orbit libraries, and 27 values of the third integral of motion over logarithmically spaced radii from 0\farcs01 to $316''$. The loop and box orbits are integrated for 2000 and 200 dynamical times, respectively.  We project the stellar orbits onto the sky and compute the LOSVDs, accounting for the KCWI point-spread function (PSF) and spatial binning. Using a non-negative least-squares optimization, we determine the orbital weights such that the linear superposition of orbits reproduces the luminous mass (to an accuracy of 1\%) and the observed kinematics in each spatial bin. As described below, the procedure is repeated for a large suite of galaxy models to determine the best combination of the galaxy model parameters. 

\subsection{Parameter search}\label{sec:Parameter_search}

The best-fit model parameters and uncertainties are determined as follows. We use an iterative grid-free Latin hypercube scheme to select sampling points in the six-dimensional model parameter space \citep{Liepoldetal2020,Quennevilleetal2022,Pilawaetal2022}. In each iteration, the \texttt{TriOS} code is run to assess the $\chi^2$ of each of the sampled galaxy models. The $\chi^2$ of a model is determined by comparing the data and uncertainties for the lowest eight kinematic moments in each of the 461 spatial bins to the model predictions.  An additional set of constraints is imposed on kinematic moments $h_9$ to $h_{12}$, in which the value of each moment is required to be zero with error bars comparable to the errors in $h_3$ to $h_8$.  As shown in \citet{Liepoldetal2020}, these additional constraints help eliminate spurious behavior in the LOSVDs predicted by the models.

The goodness-of-fit landscape is then approximated using Gaussian process regression (GPR; \citealt{RasmussenWilliamsgpml2006, scikit-learn}) with a \Matern\ covariance kernel. To map the high-likelihood region in finer detail, we run the \texttt{TriOS} code again for a next set of models selected by uniformly sampling a zoom-in volume that lies within the \nsigma{3} confidence level for six parameters in the previous regression surface. A more accurate GPR surface is then obtained from all the available models. After multiple iterations we again use GPR to construct a smooth likelihood surface from all available models (nearly 20,000 in total). Finally, we use the dynamic nested sampler \texttt{dynesty} \citep{Speagleetal2020DYNESTY} to sample from this surface to produce Bayesian posteriors assuming a uniform prior for all parameters.

Following \cite{Quennevilleetal2022}, we search over a different set of shape parameters, $T$, $T_\mathrm{maj}$, and $T_\mathrm{min}$, instead of angles $\theta$, $\phi$, and $\psi$. Such a parameterization maps the deprojectable volume in the viewing-angle space into a unit cube in the shape-parameter space, allowing for simpler and more efficient searches. The definitions of $(T, T_\mathrm{maj}, T_\mathrm{min}$) and the relationships with $(\theta, \phi, \psi)$ are given in Section~3 of \cite{Quennevilleetal2022}.

\begin{figure*}[ht]
  \centering
\includegraphics[width=6.in]{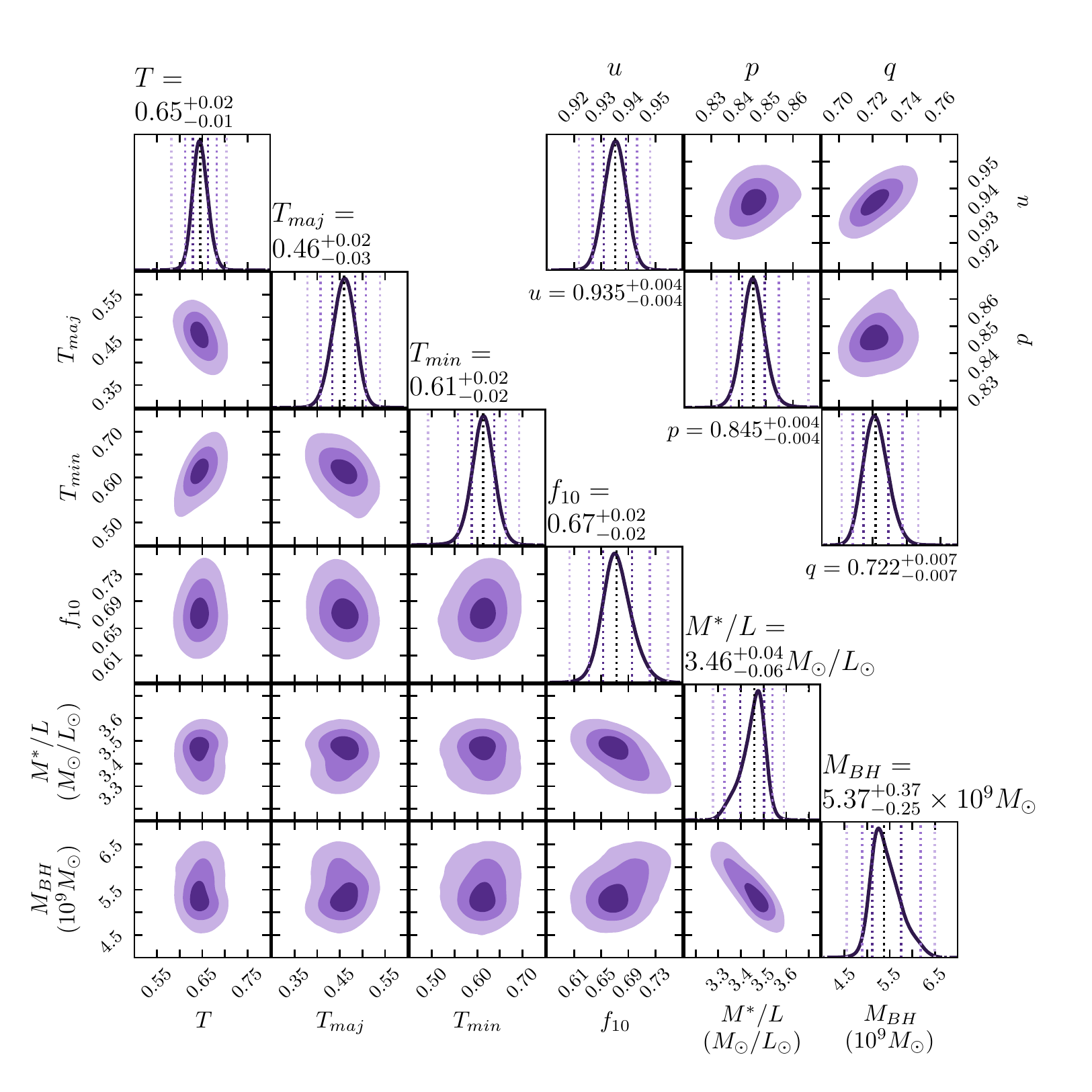}
  \caption{
  Posterior distributions of six parameters from triaxial Schwarzschild orbit modeling of M87: black hole mass \mbh, outer stellar mass-to-light ratio \ml, dark matter fraction enclosed within 10 kpc $f_{10}$, and shape parameters $T$, $T_\mathrm{maj}$, and $T_\mathrm{min}$. 
  The posterior distributions of the luminosity-averaged axis ratios $u$, $p$, and $q$ are shown in the upper right. 
  The three levels of purple shading bound the $1\sigma$, $2\sigma$, and $3\sigma$ regions (68\%, 95\%, and 99.7\% confidence levels, respectively) of the parameters. The vertical lines in each 1-dimensional distribution indicate the median and the corresponding three confidence levels. 
  }
\label{fig:Cornerplot_test}
\end{figure*}

\begin{table}
\centering
\renewcommand{\arraystretch}{1.5}
\begin{tabular}{l|l}
M87 Property (units)                                           &         Inferred value \\ \hline
Black hole mass \mbh\ ($10^9$ $M_\odot$)                       &  $5.37_{-0.25}^{+0.37} \pm 0.22$ \\ 
Outer \ml\ ($V$-band; $M_\odot / L_\odot$)                     &  $3.46_{-0.06}^{+0.04} \pm 0.15$\\
Inner \ml\ ($V$-band; $M_\odot / L_\odot$)                     &  $8.65_{-0.15}^{+0.10} \pm 0.38$ \\
Dark matter fraction at 10 kpc $f_{10}$                        &  $0.67 \pm 0.02$ \\
Total mass within 10 kpc ($10^{11} M_\odot$)                   &  $5.77 \pm 0.12$\\
Shape parameter $T$                                            &  $0.65 \pm 0.02$ \\
Shape parameter $T_\mathrm{maj}$                               &  $0.46_{-0.02}^{+0.03}$ \\
Shape parameter $T_\mathrm{min}$                               &  $0.61 \pm 0.02$ \\
Average middle-to-long axis ratio $p$                          &  $0.845 \pm 0.004$ \\ 
Average short-to-long axis ratio $q$                           &  $0.722 \pm 0.007$ \\
\multirow{1.8}{130pt}
{Average apparent-to-intrinsic long axis ratio $u$}            &  $0.935 \pm 0.004$ \\
[-2ex]                                                         &  \\
Line-of-sight direction $\theta, \phi$ $(^\circ)$              &  $48.9_{-1.0}^{+1.1}$, $37.5_{-1.3}^{+1.4}$  \\
Rotation about line-of-sight $\psi$ $(^\circ)$                 &  $-61.3_{-1.7}^{+1.4}$ \\
\end{tabular}
\caption{
Mass and shape properties of M87. The search over galaxy parameters in the triaxial orbit modeling in this paper is performed over \mbh, outer \ml, halo scale density $\rho_0$, and the shape parameters $T$, $T_\mathrm{maj}$, and $T_\mathrm{min}$. All other parameters in the table are computed from the posteriors of those six parameters. For the two primary mass parameters \mbh\ and \ml, the second set of errors denotes systematic uncertainties (68\% confidence levels) due to the uncertainty in the distance to M87: $16.8 \pm 0.7$ Mpc \citep{EHT2019f}.
}
\label{table:bf_parameters}
\renewcommand{\arraystretch}{1.0}
\end{table}

The final posterior distributions yield clear constraints on all six model parameters: \mbh, outer \ml, dark matter density $\rho_0$, $T$, $T_\mathrm{maj}$, and $T_\mathrm{min}$ (Figure~\ref{fig:Cornerplot_test}). Instead of the halo density parameter $\rho_0$, we describe the dark matter halo in terms of the ratio of dark matter to total matter enclosed within 10 kpc, $f_{10}$. The posterior distributions for the more intuitive (luminosity-averaged) axis ratios $p, q$, and $u$ are also shown. The best-fit parameters are summarized in Table~\ref{table:bf_parameters}.

\subsection{Black hole mass and stellar mass-to-light ratio} \label{sec:best_fit_parameters}

The mass of the M87 black hole has been determined with two other independent methods \citep{Walshetal2013,EHT2019a} in addition to the stellar-dynamical method used here. Compared to our value $\mbh = (5.37^{+0.37}_{-0.25}\pm 0.22) \times 10^9\ \msun$, the value $\mbh = (6.5\pm 0.2 \pm 0.7)\times 10^9\ \msun$ inferred from the crescent diameter by the Event Horizon Telescope (EHT) team \citep{EHT2019f} is 21\% higher, but the difference is within \nsigma{1.5} of their uncertainties.  A recent re-analysis of EHT observations \citep{Broderick2022} revised the black hole mass to $\mbh = (7.13\pm 0.39) \times 10^9\ \msun$, which is 33\% above our value, but \citet{Tiedeetal2022} cautioned the false-positive tendency of the method used in the re-analysis and found that significant systematic uncertainties were not taken into account. The ionized gas-dynamical determination of $\mbh = (3.45^{+0.85}_{-0.26}) \times 10^9\ \msun$ (after scaling to our adopted distance of 16.8 Mpc) is 36\% below our value \citep{Walshetal2013,EHT2019f}.

Before this work, the most recent mass measurement of the M87 black hole that also used orbit-based stellar dynamics obtained \citep{Gebhardtetal2011,EHT2019f} $\mbh = (6.14^{+1.07}_{-0.62})\times 10^9\ \msun$ (after scaling to our adopted distance of 16.8 Mpc), which is 14\% above our value. Despite the apparent consistency, there are many differences between the two measurements. In this work, the stellar spectra are obtained in a homogeneous manner from the latest IFU at the Keck Telescope over a contiguous 250 $\times$ 300\arcsec\ field and have S/N of around 100 per \AA\ for the outermost bins and above 200 per \AA\ for central bins. The observed stellar velocity dispersions used to constrain the orbit models in this work are about 20\% lower than \citet{Gebhardtetal2011,Murphyetal2011} beyond 1 kpc (Figure~\ref{fig:dispersion}; top panel), but this work is in broad agreement with other recent measurements \citep{Emsellemetal2014,Sarzietal2018,Forbesetal2020}. The orbit modeling in this work allows for triaxiality, and the \mbh\ is obtained from a full six-dimensional model parameter search with posteriors measured using a Bayesian framework. Furthermore, \citet{Gebhardtetal2011} adopts a spatially constant $V$-band \ml\ of $9.7\ M_\odot/L_\odot$ (scaled to our distance of 16.8 Mpc). However, a recent detailed stellar population analysis of M87 reported a negative radial gradient due to a changing stellar initial mass function \citep{Sarzietal2018}. When incorporating the shape of this \ml\ gradient into our stellar-dynamical models, we find the $V$-band \ml\ declines from $8.65\ M_\odot/L_\odot$ at the center to an outer value of $3.46\ M_\odot/L_\odot$.

Using either $M^*(< r_{\rm SOI})=\mbh$ or $M^*(< r_{\rm SOI})=2\mbh$ as the definition of a black hole's gravitational sphere of influence (SOI), we find the SOI radius of the M87 SMBH to be $r_{\rm SOI}=4\farcs4$ (0.36 kpc) or 6\farcs1 (0.50 kpc).

\subsection{Dark matter mass}

At the outer reach of our data, at a radius of 10 kpc, we find the enclosed dark matter mass to be $M_{\rm DM}(<10\,{\rm kpc})=(3.88\pm 0.12)\times 10^{11}\ \msun$, constituting about 67\% of the total mass of the galaxy ($f_{10}$ in Table~\ref{table:bf_parameters}). A similar dark matter fraction (73\% at 14.2 kpc) is obtained from Jeans modeling of the kinematics of globular clusters \citep{Lietal2020}. A lower dark matter fraction (about 30\% at 11 kpc) is estimated from axisymmetric orbit-based modeling of the kinematics from stars and globular clusters \citep{Murphyetal2011}. This lower fraction arises mainly from their high estimate of \ml\ discussed in the previous paragraph.

Our inferred total mass of M87 within 10 kpc is $M_{\rm tot}(< 10\,{\rm kpc})=(5.77\pm 0.12)\times 10^{11}\ \msun$. Dynamical modeling of globular clusters under the assumption of spherical symmetry yields very similar value at the same radius \citep{WuTremaine2006,RomanowskyKochanek2001} but with large modeling uncertainties \citep{WuTremaine2006}. Estimates from axisymmetric orbit models find a 15\% lower value \citep{Murphyetal2011}. Jeans modeling studies \citep{OldhamAuger2016,Lietal2020} incorporating a radially declining \ml\ find a total mass enclosed within 10 kpc to be in the range of $(3$\textendash$7.5)\times 10^{11}\ \msun$.

\subsection{M87's intrinsic shape} \label{sec:intrinsic_shape}

Our orbit modeling results show that M87 is strongly triaxial, where the lengths of the short and middle principal axes are 72\% and 85\% of the length of the long axis, corresponding to $q$ and $p$, respectively. A triaxiality parameter often used to quantity the ratios of a galaxy's principal axes is $T = (1 - p^2)/(1 - q^2) = (a^2 - b^2)/(a^2-c^2)$. This parameter ranges between $T=0$ for an oblate axisymmetric shape ($p=1$ or $a=b$) and $T=1$ for a prolate axisymmetric shape ($p=q$ or $b =c$), with values between 0 and 1 indicating a triaxial shape.  Our inferred value for M87 is $T = 0.65 \pm 0.02$, strongly excluding the possibility that M87 is an axisymmetric galaxy.

The shape parameters $p$, $q$, and $u$ in Table~\ref{table:bf_parameters} are related to a set of angles $\theta$, $\phi$, and $\psi$ that uniquely specify the orientation of M87's intrinsic axes with respect to its projected axes on the sky \citep{vandenBoschetal2008,Quennevilleetal2022}. The angles $\theta$ and $\phi$ specify the direction of the line-of-sight from M87 to the observer; they are the usual polar angles in M87's intrinsic coordinate system. The inclination angle $\theta=0^\circ$ corresponds to a face-on view of M87 along its intrinsic short axis, and $\theta=90^\circ$ corresponds to an edge-on view with the short axis in the sky plane. The azimuthal angle $\phi = 0^\circ$ places the intrinsic middle axis in the sky plane and $\phi = 90^\circ$ places the intrinsic long axis in the sky plane. Once the line of sight is described by $\theta$ and $\phi$, the third angle $\psi$ specifies the remaining degree of freedom for the rotation about the line of sight. Our best-fit angles for M87 are $(\theta, \phi, \psi) = (48\fdg9, 37\fdg5, -61\fdg3)$. Thus, we are viewing M87 from a direction that is roughly equidistant from all three principal axes. 

\subsection{Angular momentum vector and origin of kinematic misalignment}

To gain physical insight into the origin of the observed misalignment between the kinematic axis and photometric major axis of M87 on the sky (Figure~\ref{fig:velocity}; lower right), we examine the direction of the total angular momentum vector, $\vect{L}$, of the stars predicted by our best-fit orbit model and how it would be projected on the sky. To do this, we sum the individual contributions to the angular momentum from the superposition of stellar orbits and compute the total $\vect{L}$. Among the three major orbital types computed in the \texttt{TriOS} code, the box orbits supported by a triaxial gravitational potential, by construction, have zero angular momentum, whereas the short-axis and long-axis tube orbits have net $\vect{L}$ along the intrinsic short axis and long axis, respectively \citep{Schwarzschild1979,vandenBoschetal2008,Quennevilleetal2022}. The direction of the total $\vect{L}$ is therefore determined by the relative contributions from the two types of tube orbits \citep{Franxetal1991}.

The rotational velocity of M87 reaches sufficiently high amplitudes beyond about 5 kpc (Figures~\ref{fig:kin_map} and \ref{fig:velocity}) for us to determine the direction of $\vect{L}$ robustly. We find it to point approximately $60^\circ$ off of the intrinsic short axis. Using the best-fit viewing angles to project $\vect{L}$ on the sky, we find it to lie at a PA of approximately $-60^\circ$. Because the projected $\vect{L}$ is orthogonal  to the kinematic axis of the projected velocity field, this simple calculation indicates that the PA of the kinematic axis predicted by the model is around $-150^\circ$, very similar to the observed kinematic axis. The observed kinematic misalignment of M87 on the sky is therefore a result of both projection effects of a triaxial galaxy and a physical offset between the total angular momentum vector and the intrinsic short axis of the galaxy.

\section{Conclusions}

With 4000 constraints from Keck KCWI and our latest triaxial orbit modeling code and procedure for sampling high-dimensional parameter spaces even with computationally intensive models, we are able to relax the common assumption of axisymmetry and present the most comprehensive stellar-dynamical study of the M87 galaxy and its central black hole. This work is one of only a small number of studies that have produced constraints on all three intrinsic shape parameters for individual galaxies \citep{Jinetal2020, Santucci2022}. Even fewer galaxies have been observed with sufficient angular resolution, field of view, spectral coverage, and S/N for a simultaneous determination of the intrinsic shape, supermassive black hole mass, and galaxy mass \citep{vdb10,Walshetal2012,denBroketal2021, Quennevilleetal2022,Pilawaetal2022}. As demonstrated in this work, further advancements have only been made possible by the installations of wide-field and highly sensitive IFUs on large ground-based telescopes.  

Moving forward, it is crucial to apply triaxial stellar-dynamical orbit models to larger samples of galaxies, thereby advancing this method from a rarity to a standard technique. This is especially pertinent for massive elliptical galaxies such as M87 because the majority of them\textemdash when a rotational pattern can be detected in the stellar velocity field\textemdash show some degree of misalignment between the kinematic and photometric major axes, extending to the half-light radius and beyond \citep{Eneetal2018,Krajnovicetal2018,Eneetal2020}. Such an offset indicates triaxiality \citep{Binney1985,Franxetal1991}; an axisymmetric galaxy would, by symmetry, produce only aligned kinematic and photometric major axes.

When direct comparisons between axisymmetric and triaxial modeling were made on the same galaxy, the black hole mass from axisymmetric models has ranged from about 50\% \citep{vdb10} to 170\% \citep{Pilawaetal2022} of the mass when triaxiality was allowed; and in two galaxies, the black hole mass did not change appreciably \citep{vdb10, Liepoldetal2020,Quennevilleetal2022}.  Overall, triaxial models were able to match the observed stellar kinematics significantly better than axisymmetric models \citep{Quennevilleetal2022,Pilawaetal2022}. 
 
More secure black hole masses could result in significant changes to the local black hole census and the shapes of the scaling relations between black holes and host galaxies, thereby impacting our understanding of black hole fueling and feedback physics, as well as binary black hole merger physics used to forecast and eventually interpret gravitational wave signals for Pulsar Timing Arrays \citep{taylor2021} and space-based detectors \citep{LISA2022}. 
In terms of black hole imaging studies, since the photon ring diameter ranges from about 9.6 to 10.4 gravitational radii depending on the black hole spin \citep{EHT2019f}, future analyses combining direct imaging with stellar kinematic measurements such as that presented this paper have the potential to significantly improve the prospects for measuring black hole spins.

\section*{Acknowledgments}
We dedicate this work to the late Wal Sargent, who reported the first observational evidence for the M87 black hole and was a mentor to generations of scientists including C.-P.M. We thank Scott Tremaine, Michael Johnson, Charles Gammie, and the referee for insightful comments.
E.R.L. and C.-P.M. are supported by NSF AST-1817100 and AST-2206307.  J.L.W. is supported by NSF AST-1814799 and AST-2206219. C.-P.M. acknowledges the support of the Heising-Simons Foundation and the Miller Institute for Basic Research in Science. The spectroscopic data presented in this paper were obtained at the W. M. Keck Observatory, which is operated as a scientific partnership among the California Institute of Technology, the University of California and the National Aeronautics and Space Administration. The Observatory was made possible by the generous financial support of the W. M. Keck Foundation. 
This work used observations made with the NASA/ESA Hubble Space Telescope, obtained at the Space Telescope Science Institute, which is operated by the Association of Universities for Research in Astronomy, Inc., under NASA contract NAS5-26555. This work used the Extreme Science and Engineering Discovery Environment (XSEDE) at the San Diego Supercomputing Center through allocation AST180041, which is supported by NSF grant ACI-1548562.\\[2ex] 

\vspace{5mm}
\facilities{Keck (KCWI)}

\software{
\texttt{TriOS} \citep{Quennevilleetal2021,Quennevilleetal2022},
orbit code \citep{vandenBoschetal2008},
\texttt{dynesty} \citep{Speagleetal2020DYNESTY},
\texttt{scikit-learn} \citep{scikit-learn},
\texttt{LHSMDU} \citep{deutschdeutsch2012,LHSMDU},
\texttt{pPXF} \citep{Cappellari2017},
\texttt{vorbin} \citep{CappellariCopin2003},
\texttt{MGE} \citep{Cappellari2002},
KCWI Data Reduction Pipeline \citep{Morrisseyetal2018}
}

\appendix

\section{Keck KCWI data reduction and analysis \label{appendix:data_analysis}}
\renewcommand{\thefigure}{\ref{appendix:data_analysis}\arabic{figure}}
\setcounter{figure}{0}

We observed M87 using the integral-field spectrograph KCWI on Keck. We used the BL grating centered on $4600$ \AA\ and the Kblue filter to obtain the widest wavelength coverage, reducing possible template mismatch during the subsequent extraction of the stellar kinematics. The integration time per exposure varied from 300 s for the central pointings to 1500 s for the outermost pointings with low surface brightness. We periodically acquired offset sky exposures in between the on-source galaxy exposures, each roughly half the integration time of the adjacent galaxy exposures.  Only data taken in good observing conditions are used in this analysis; the on-source and sky exposure times total 13 hr and 2.8 hr, respectively.

\subsection{Data reduction}

The KCWI Data Extraction and Reduction Pipeline \citep{Morrisseyetal2018} is actively  maintained on a publicly accessible GitHub repository. We use the \texttt{IDL} version of the pipeline with its default settings to reduce each frame. The main steps include overscan and bias removal, cosmic ray rejection, dark and scattered light subtraction, solving for the geometric distortion and wavelength solution, flat-fielding, correction for vignetting and the illumination pattern, sky subtraction, and the generation of datacubes using the spatial and spectral mappings determined previously. The pipeline then corrects for differential atmospheric refraction and applies a flux calibration using a standard star.

In addition to the default pipeline, we perform custom steps to improve the quality of the processed data. Some cosmic rays are improperly removed by the KCWI pipeline, leaving sharp features at certain wavelengths in a small number of spaxels in our datacubes. We therefore scan through each wavelength slice of the cubes, mask the impacted pixels, and perform an interpolation to replace their values with those of neighboring pixels. Furthermore, beyond about 100\arcsec, the KCWI spectra are sky-dominated and subtle mis-subtraction of the sky can result in significant reduction of the S/N of the galaxy spectra. The sky subtraction stage of the KCWI pipeline uses b-spline interpolation to build a ``noise-free'' model of the sky in each pixel that is subtracted from the corresponding object exposure. We find that this routine does not capture highly space- or time-variant sky features, so we further remove residual sky features using the combination of a principal component analysis (PCA) and the penalized pixel-fitting (pPXF; \citealt{Cappellari2017}) method, as described in Appendices \ref{appendix:pca} and \ref{appendix:kin_measurement}.

In the final step, we merge the on-source M87 datacubes. Roughly half of the pointings were taken with the long axis of KCWI aligned with a PA of $-25^\circ$ and half were oriented perpendicular to this with a PA of $-115^\circ$. We construct a pair of datacubes, one for each of the two orientations using the \texttt{nifcube} and \texttt{gemcube} IRAF tasks that are part of Gemini's data reduction software. We input the fully calibrated KCWI datacubes (the ``\_icubes.fits'' files) and map the cubes onto a shared grid with a spacing of 0\farcs3 $\times$ 1\farcs4 $\times$ 1 \AA. This choice of spaxel size matches the native scale of the individual KCWI datacubes for our observational setup.

\subsection{Line-spread function}\label{sec:appendix_LSF}

\begin{figure}
\centering
\includegraphics[width=3.in]{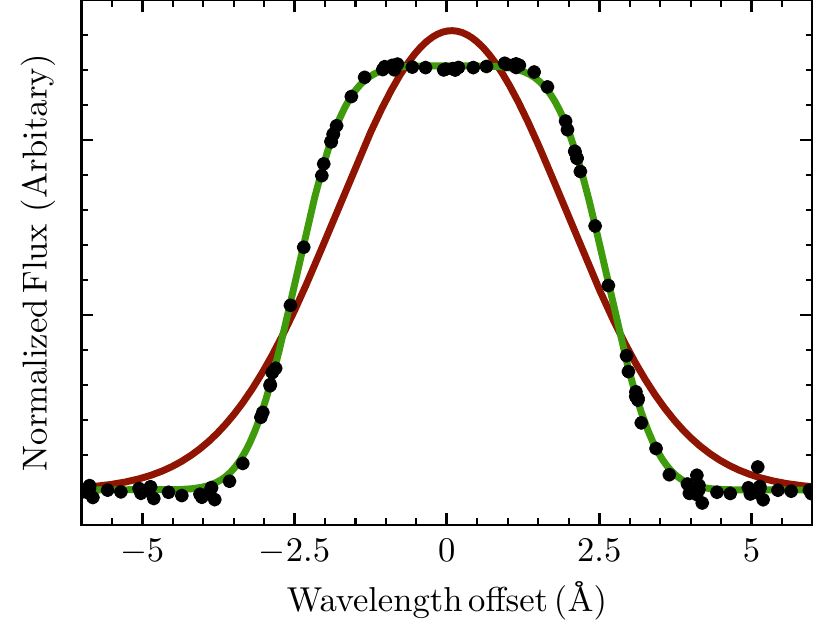}
\caption{
Line Spread Function of Keck KCWI with BL grating.
We find the LSF of KCWI BL grating to be well approximated by a Gaussian function convolved with a top-hat function, as shown in Equation \ref{eqn:LSF}. To measure the shape of the LSF, we simultaneously fit 31 lines of an FeAr lamp spectrum as described in Appendix~\ref{sec:appendix_LSF}. Here we plot a superposition of the nine most prominent of those lines. Black points mark the flux in the lamp spectrum around each line after normalizing for each line's amplitude. Our best-fit LSF model (green) has a top-hat function of width $\Delta=5.105$ \AA\ convolved with a Gaussian function of $\sigma=0.627$ \AA. A single Gaussian function, as is typically assumed, would  provide a very poor fit to the KCWI LSF (red).
}
\label{fig:LSF}
\end{figure}

We find that our selected spectrograph configuration produces a line-spread function (LSF) that is distinctly non-Gaussian (Figure~\ref{fig:LSF}). The LSF is instead well described by the convolution of a Gaussian function and a top-hat function of the form
\begin{equation}
\label{eqn:LSF}
\begin{split}
\mathcal{L}(\lambda) &= \int_{-\infty}^\infty \frac{1}{\sqrt{2 \pi \sigma^2}} e^{-\frac{(\lambda - \tau)^2}{2 \sigma^2}} \Pi \left( \frac{\tau}{\Delta}\right)\, d\tau\\
&= \frac{1}{2} \left[ \mathrm{erf} \left(\frac{\lambda + \Delta/2}{\sigma}\right) - \mathrm{erf} \left(\frac{\lambda - \Delta / 2 }{\sigma} \right) \right],
\end{split}
\end{equation}
where $\Pi(x) = 1$ if $|x| \le 1/2$ and $0$ otherwise, $\Delta$ is the full width of the top-hat component, and $\sigma$ is the standard deviation of the Gaussian. To measure the widths of the Gaussian and top-hat components of the LSF, we simultaneously fit 31 lines of an FeAr arc lamp spectrum between 4500 and 5000 \AA\ and determine $\Delta = \SI{5.105}{\angstrom}$ and $\sigma = \SI{0.627}{\angstrom}$. Repeating this procedure on different spectral or spatial regions yields comparable best-fit parameters.

\subsection{Point-spread function}

During the first night of observations, we took KCWI data of the inner region of M87 and the atmospheric seeing was estimated to be 0\farcs63 by the differential image motion monitor at the nearby Canada France Hawaii Telescope weather station. This estimate is consistent with the broadening of point sources measured from exposures taken with the guider camera.  During the other four nights, we observed the outer regions of M87 and measured similar seeing. While running stellar-dynamical models, described in Section \ref{sec:orbit_modelling}, we use a PSF that is a Gaussian with a full width at half maximum (FWHM) of 0\farcs63 ($\sigma = 0\farcs28$).

\subsection{PCA decomposition of sky features \label{appendix:pca}}

As part of the process to remove residual sky features seen in the reduced M87 datacubes, we perform a PCA decomposition of the sky spectra.  For each sky cube, we apply a conservative spatial masking of possible sources in the field and coadd the unmasked pixels to obtain a high S/N sky spectrum. A weighted expectation-maximization PCA \citep{Bailey2012} is then applied to each of the sky spectra between 3800 and 5650 \AA. Since the amplitudes of the 4861 \AA\ H$\beta$, 5200 \AA\ [N I], and the 5577 \AA\ [O I] lines are highly variable and are not well captured with a PCA decomposition \citep{Vandokuumetal2019}, we mask these features. The first PCA component is effectively the mean sky spectrum. The second and fourth components capture slight variations in the shape of the continuum and the Ca H and K features. The third and fifth components capture variations in the numerous OH lines. While we obtain measurements of the first ten components, the fifth component and beyond are consistent with noise. A similar routine was previously applied to KCWI observations \citep{Vandokuumetal2019} and this method is similar in spirit to the Zurich Atmospheric Purge (ZAP; \citealt{Sotoetal2016}) used for MUSE observations.

\subsection{Spectral and spatial masking}

\begin{figure}
\centering
\includegraphics[width=3.2in]{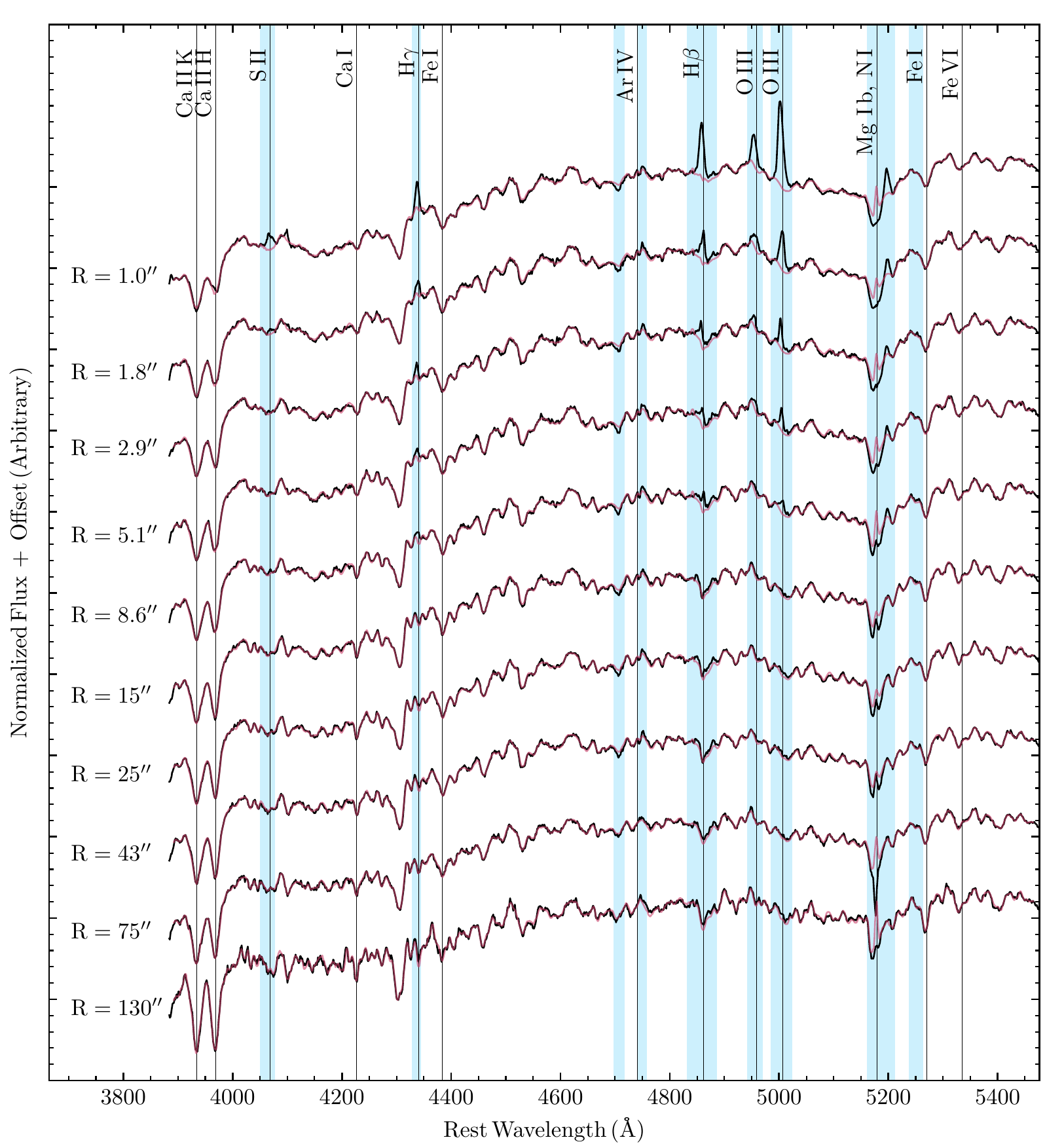}
\hfill
\caption{
Representative KCWI spectra of M87.
Sky-subtracted galaxy spectra (black curves) for ten representative spatial bins located at projected radii from 1\arcsec\ to 130\arcsec\ are shown. A total of 461 binned spectra are used in this work. The S/N of these co-added spectra range from about 100 to 200 per \AA.  The stellar template broadened by the best-fit LOSVD is overlaid (red curves) on each spectrum. Typical fitting residuals are comparable to the line widths. Shaded blue regions indicate masked spectral regions excluded from the analysis. At M87's redshift, the \mgb\ region is contaminated by the 5200 \AA\ [N I] sky line (marked). The central bins exhibit strong AGN emission lines, especially from [OIII] (4959 \AA\ and 5007 \AA), [SII] (4069 \AA\ and 4076 \AA), H$\beta$ (4861 \AA), and H$\gamma$ (4330 \AA).
}
\label{fig:spectra}
\end{figure}

We mask nine spectral features, which together span a total of 274 \AA\ (Figure~\ref{fig:spectra}). The masked features include emission lines that are prominent at the nucleus, as well as the 4861 \AA\ H$\beta$, 5200 \AA\ [N I], and 5577 \AA\ [O I] lines that are masked in the PCA decomposition.  The \mgb\ region ($5184$\textendash$5234$ \AA) is also masked because it is contaminated by the \SI{5200} \AA\ [N I] skyline and is coincident with Fe emission features at M87's redshift.

We also apply a spatial mask to exclude potentially contaminant spaxels. This is done by collapsing the datacubes spectrally, flagging regions of spaxels with substantially higher surface brightness than their surroundings, and then masking the brightest spaxels in those regions. This process removes the spaxels that are contaminated by the prominent jet, the central ${\sim}0\farcs85$ that is affected by the active galactic nucleus (AGN), and numerous bright globular clusters.

\subsection{Spatial binning} 

We use the \texttt{vorbin} package \citep{CappellariCopin2003} to construct spatial bins and obtain coadded KCWI stellar spectra with uniformly high S/N. By default, \texttt{vorbin} calculates the S/N of each coadded spectrum based on values of the signal and the noise of the individual spaxel spectrum given by the user, adding the signals linearly and the noise in quadrature. Instead of this default setting, we modify the \texttt{sn\_func()} routine in \texttt{vorbin}'s \texttt{voronoi\_2d\_binning} to nonanalytically recompute the S/N from the M87 datacube while binning. This approach improves the uniformity of the resultant S/N across the bins as it naturally incorporates spatial correlations in the signal and noise between spaxels. We estimate the S/N by first smoothing the spectrum with a Gaussian kernel with FWHM = 4 \AA, comparable to the LSF. The noise is taken to be the root-mean-square (rms) difference between the raw and smoothed spectra, while the signal is taken to be the median flux of the raw spectrum. We apply the spectral masks described above before smoothing to avoid contamination from sharp features in the spectra.

This procedure results in 461 spatial bins and a coadded spectrum for each of the bins. The S/N per \AA\ ranges from about 200 in the central regions to about 100 in the outer regions. Figure~\ref{fig:spectra} shows a series of representative KCWI spectra (black curves) for ten of the 461 spatial bins located at projected radii of 1\arcsec\textendash 130\arcsec. 

\section{Stellar kinematic determination \label{appendix:kin_measurement}}
\renewcommand{\thefigure}{\ref{appendix:kin_measurement}\arabic{figure}}
\setcounter{figure}{0}

We measure the stellar LOSVD for each of the 461 binned spectra using pPXF \citep{Cappellari2017}. With pPXF, we convolve a linear combination of template stars with an LOSVD, parameterized by $V$, $\sigma$, and high-order Gauss-Hermite moments $h_3$\textendash$h_8$\ that account for asymmetric and symmetric deviations from a Gaussian velocity distribution \citep{vanderMarelFranx1993}. The S/N of our data enable the measurement of high-order Gauss-Hermite moments. We find that truncating the series at $h_4$ (or $h_6$) results in elevated values for $h_4$ (or $h_6$), but when fitting to $h_8$ or $h_{12}$, the values of $h_4$ and $h_6$ converge and the highest extracted moments become consistent with $0$, as seen in past work \citep{Pilawaetal2022,Liepoldetal2020}. In addition, we find it important to constrain the kinematic moments beyond $h_4$ in dynamical modeling. When those moments are not constrained in orbit models, the models are prone to producing LOSVDs with unphysical features due to large values in the high-order moments, potentially biasing the preferred model parameters \citep{Quennevilleetal2021,Liepoldetal2020}.

For stellar templates, we use the MILES library \citep{Falcon-Barrosoetal2011, sanchezetal2006} but select 485 spectra out of the full 985 templates that have well-identified spectral types and luminosity classifications. These stellar templates have a higher spectral resolution than our observations and are degraded to match the KCWI (non-Gaussian) LSF before fitting with pPXF.

During the kinematic fit, we use an additive polynomial of degree one and a multiplicative polynomial of degree 15 to model the stellar continuum. We also supply the PCA components that describe the sky background to pPXF. This procedure results in a weighted combination of the PCA components, which is included as an additional additive term to match the residual sky features in the M87 spectra that remained after the KCWI pipeline's default sky subtraction. Ultimately, we use the first ten PCA components, but find that the extracted Gauss-Hermite moments are unchanged as long as at least the first five PCA components are included in the fit. 

Because we excluded three highly variable sky lines during the PCA decomposition process, we also mask those spectral regions when running pPXF, as well as emission lines associated with M87 and the \mgb\ region, as described previously. In contrast to the other masked regions, we find that the extracted Gauss-Hermite moments depend strongly on the endpoints of \mgb\ mask and only stabilize once the entire $5184$\textendash$5234$ \AA\ region is excluded from the fit. 

Altogether, we fit the Gauss–Hermite moments, polynomial coefficients, template weights, and sky weights simultaneously. The stellar templates broadened by the best-fit LOSVD provide excellent fits to each of the observed spectra, as illustrated by the red curves for the ten representative spectra shown in Figure~\ref{fig:spectra}.

The measurement uncertainties on the LOSVDs are determined as follows. After an initial fit to each binned spectrum, we perturb the spectrum at a given wavelength by drawing a random number from a Gaussian distribution centered on the spectrum and with a dispersion equal to the rms of the pPXF residuals from the preliminary fit at that wavelength. We perform 1000 such perturbed fits with the pPXF bias parameter set to $0$ and determine the mean and standard deviation of each moment over those 1000 realizations, which we adopt as the kinematic value and its \nsigma{1} uncertainty. For bins in the central 100\arcsec\ $\times$100\arcsec\ region, the mean error on $V$ is $2.6\ \kms$ and on $\sigma$ is $3.0\ \kms$. The mean errors on $h_3$ through $h_8$ are similar, spanning from 0.009 to 0.016. The typical errors in the outer bins are slightly larger with mean errors on $V$, $\sigma$, and $h_3$ through $h_8$ of $2.6\ \kms$, $3.4\ \kms$, and $0.012$\textendash$0.022$, respectively.

\begin{table}
\begin{tabular}{c|c|c}
\hline
$I_k \,[L_{\odot} / {\rm pc}^{2}]$ & $ \ \ \sigma_k^{\prime} \ \ [\mathrm{arcsec}] \ \ $ & $ \ \ q_k^{\prime} \ \ $ \\
\hline
        $2382.4$   &   $0.039$   &        $0.860$  \\
        $2460.8$   &   $0.206$   &         $0.906$  \\
        $1598.8$   &   $0.508$   &         $0.959$  \\
        $973.48$   &   $1.468$   &          $1.000$  \\
        $1830.9$   &   $4.558$   &          $1.000$  \\
        $1515.4$   &   $9.851$   &          $0.980$  \\
        $592.67$   &   $22.228$   &          $0.942$  \\
        $180.27$   &   $54.299$   &           $0.910$  \\
        $35.555$   &   $126.213$   &          $0.735$  \\
        $9.8327$   &   $293.034$   &          $0.650$  \\
        $1.6948$   &   $567.511$   &          $1.000$  \\
\end{tabular}
\centering
\caption{
Best-fit MGE parameters for the surface brightness of M87.
For each of the 11 two-dimensional Gaussian components, the first column lists the central surface brightness density, the middle column lists the dispersion of the Gaussian, and the last column lists the axis ratio, where primed variables denote projected quantities. We obtain the MGE by fitting to the $V$-band light profile in \citet{kormendyetal2009}. To impose a $\ml$ gradient in the dynamical models, the $I_k$ values are adjusted to reproduce the profile in Figure~\ref{fig:ML_gradient}.
}
\label{tab:mge}
\end{table}

\section{Surface brightness of M87 \label{appendix:mge}}
\renewcommand{\thefigure}{\ref{appendix:mge}\arabic{figure}}
\setcounter{figure}{0}

Besides the stellar kinematics, another constraint used in the dynamical models is the galaxy's luminosity density. We use a previously published $V$-band light profile, along with measurements of the ellipticity and PA of the isophotes \citep{kormendyetal2009}. The profile extends from 0\farcs017 to 2400\arcsec\ and comes from a combination of ground-based data and high-resolution HST images, which have been deconvolved to remove the effects of the PSF as well as the AGN.

We fit the sum of multiple two-dimensional Gaussians to the composite surface photometry. These MGE \citep{Cappellari2002} approximations are commonly used because they are able to match the surface brightnesses of galaxies while also enabling analytical deprojections to obtain intrinsic luminosity densities.  Our best-fit MGE reproduces the surface brightness between 0\farcs1 and 500\arcsec\ within 10\%. This MGE has 11 Gaussian components that share the same center and PA of $-25^\circ$.  While the value of the PA in \citet{kormendyetal2009} varies within 50\arcsec, the isophotes between 1\arcsec\ and 50\arcsec\ are very round with ellipticity $\epsilon \lesssim 0.08$; using a constant PA in our MGE therefore does not affect the quality of the fit. The MGE parameters are given in Table~\ref{tab:mge}.

\section{Orbit modeling}\label{sec:appendix_orbit_modelling}
\renewcommand{\thefigure}{\ref{sec:appendix_orbit_modelling}\arabic{figure}}
\setcounter{figure}{0}

\begin{figure}[t]
\centering
\hfill
\includegraphics[width=3.25in]{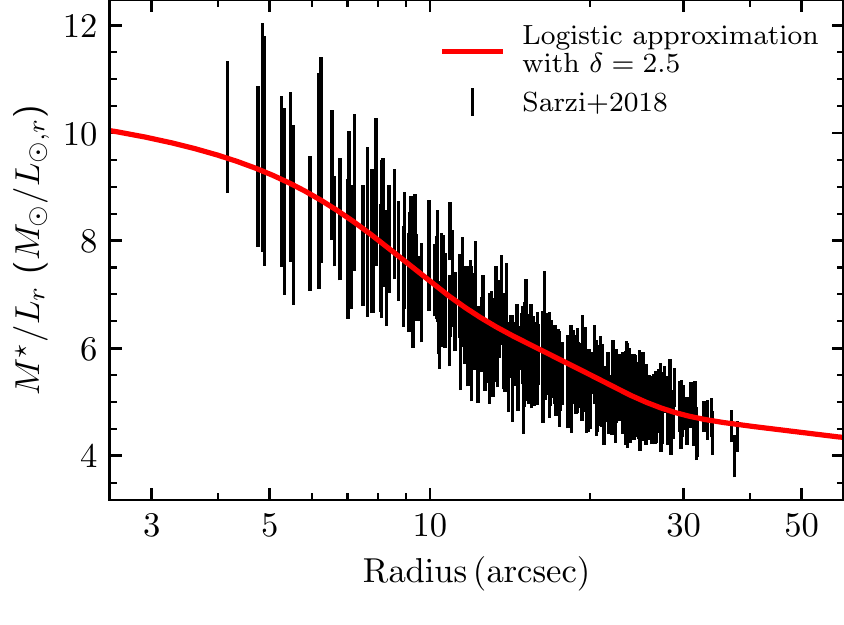}
\hfill
\caption{
Radial profile of \ml\ ratio used in this work.
The logistic approximation (red) used in our modeling, given by Equation~(\ref{eqn:MLgradient}), is chosen to match the shape of the $r$-band \ml\ (black) in Figure 11 of \citet{Sarzietal2018}. The inner \ml\ is $\delta=2.5$ times the outer \ml\ ratio, and the transition is centered around 10 arcsec. Our dynamical model prefers an outer $V$-band \ml\ of $3.46_{-0.06}^{+0.04} M_\odot/L_\odot$ and inner \ml\ of $8.65_{-0.15}^{+0.10}$ $M_\odot/L_\odot$.
}
\label{fig:ML_gradient}
\end{figure}

\begin{figure}[!t]
\centering
\includegraphics[width=3.25in]{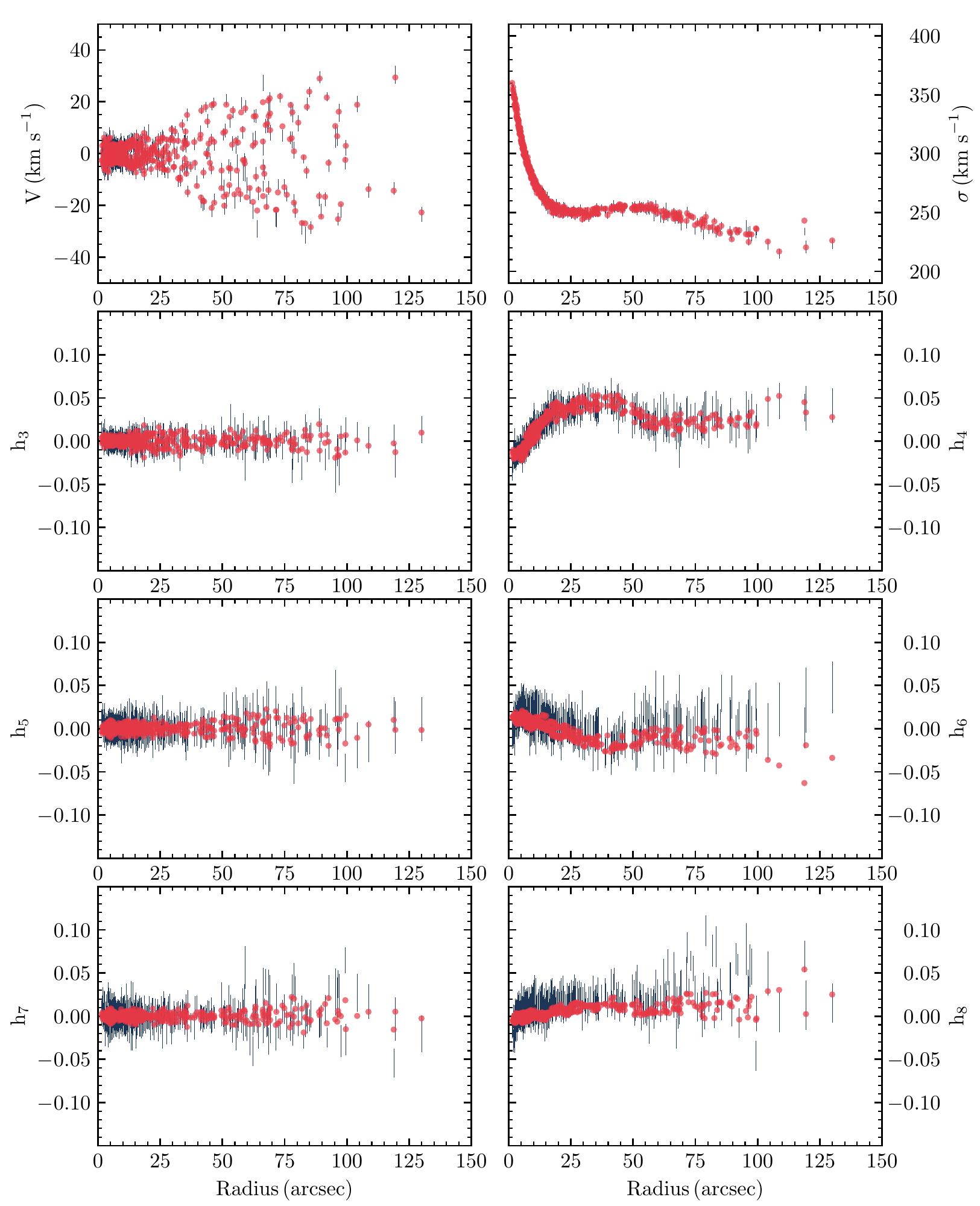}
\hfill
\caption{
Radial profiles of the first eight moments of the stellar LOSVDs.
The observed Keck KCWI moments (gray) are well matched by the moments predicted by the best-fit model (red) given by Table~\ref{table:bf_parameters}. The triaxial orbit models produce point-symmetric LOSVDs, so we have point-symmetrized the kinematic moments before fitting.
}
\label{fig:kinematics}
\end{figure}

Radial profiles of the \ml\ ratio and stellar kinematics used in the orbit modeling in this work are shown in Figure~\ref{fig:ML_gradient} and Figure~\ref{fig:kinematics}, respectively.

\clearpage

\bibliography{M87}{}
\bibliographystyle{aasjournal}

\end{document}